\documentclass[aps,pr,reprint,groupedaddress,floatfix]{revtex4}
\usepackage{amsmath,amssymb,graphicx}

\begin{document}

\title{Application of a multi-site mean-field theory\\ to the disordered Bose-Hubbard model}

\author{P. Pisarski}
\affiliation{Department of Physics, Engineering Physics, and Astronomy, Queen's University, Kingston, ON K7L 3N6, Canada}

\author{R.M. Jones}
\affiliation{Department of Physics, Engineering Physics, and Astronomy, Queen's University, Kingston, ON K7L 3N6, Canada}

\author{R.J. Gooding}
\affiliation{Department of Physics, Engineering Physics, and Astronomy, Queen's University, Kingston, ON K7L 3N6, Canada}

\date{\today}

\begin{abstract}
We present a multi-site formulation of mean-field theory applied to the disordered Bose-Hubbard model.
In this approach the lattice is partitioned into clusters, each isolated cluster being treated exactly,
with inter-cluster hopping being treated approximately.
The theory allows for the possibility of a different superfluid order parameter at every site in the lattice, such as
what has been used in previously published site-decoupled mean-field theories, 
but a multi-site formulation also allows for the
inclusion of spatial correlations allowing us, \textit{e.g.}, to calculate the correlation length (over the length scale
of each cluster). We present our numerical results for a two-dimensional system.
This theory is shown to produce a phase diagram in which the stability of the Mott insulator phase is
larger than that predicted by site-decoupled single-site mean-field theory.  
Two different methods are given for the identification of the Bose glass-to-superfluid transition, one an approximation
based on the behaviour of the condensate fraction, and 
one of which relies on obtaining the spatial variation of the order parameter correlation.
The relation of our results to a recent proposal that both transitions are non self-averaging is discussed.
\end{abstract}

\pacs{
PACS numbers:67.85.Hj
}

\maketitle
 \section{\label{sec:intro}Introduction}

The experimental realization of a gas of ultracold atoms in an optical lattice \cite{Jaksch} has provided a unique opportunity to
exploit a remarkably close connection between experimental and theoretical studies of interacting Bose condensed systems.
In one sense, one may think of the experiments as constituting a ``quantum simulator" of strongly interacting particles on a lattice \cite{qsimulator} described by the Hubbard model \cite{Hubbard}. With the subsequent addition of speckle-generated disorder \cite{speckle} to these experiments, or by the generation of multichromatic lattices \cite{multichromatic}, a window into the world of strongly interacting disordered physics was opened.

The seminal paper of Fisher, \textit{et al.} \cite{Fisher89}, in 1989 predates the production of Bose-condensed ``cold" atomic gases 
by several years. The focus of that paper is the disordered Bose-Hubbard model, including the thermodynamically stable quantum phases, the properties of these phases, and the transitions between them. Many theory papers have followed since this work, the number of such papers is growing (nearly) exponentially since the realization of this model via cold atom experiments. 

One of the simplest and most often employed methods to analyze a model such as the disordered Bose-Hubbard model is
mean-field theory. The paper of Fisher, \textit{et al.} \cite{Fisher89}, used such a theory to provide the first qualitative phase
diagram of this system. Many mean-field theory investigations, some of which we now mention, have followed this paper. Our contribution, presented in this report, is the application of a so-called multi-site mean-field theory to investigate some of the same questions addressed in Ref. \cite{Fisher89}, with the ultimate goal of making comparisons between such
a theory and experiment. As we outline below, the hoped-for improvement of such a multi-site theory is the incorporation
of some spatial correlations that are neglected in conventional single-site site-decoupled mean-field analyses.
 
As stated above, the present work is an example of a multi-site mean-field theory, and to put this approach into perspective it is important to first discuss
the set of results found from the familiar site-decoupled single-site mean-field theory. 
The original paper of Fisher, \textit{et al.} \cite{Fisher89}, provided the qualitative aspects of the phase boundaries separating the relevant quantum
phases, namely the Mott insulating, Bose glass, and superfluid phases. Some time ago, using an RPA-based approach,
this phase diagram was obtained quantitatively \cite{Ramakrishnan}. Following the many experimental
advances discussed above, this theory was re-examined by several groups. We mention first the spatially-averaged
single-site mean-field theory of Krutitsky, \textit{et al.} \cite{Krutitsky}. As discussed in the next section, this formulation allows one to 
obtain the phase boundaries analytically. However, since one is treating each site as being equivalent to all other sites, 
the resulting theory does not include any of the effects of coherent back-scattering that give rise to localized states, and
correlations between the order parameter at neighbouring sites are completely ignored (see Eq.~(\protect\ref{eq:phiiphij}) below). 
This limitation may be overcome by allowing for a different order parameter at every site, something that was achieved in a series of detailed papers
(below we refer specifically to only three papers in this series, Refs. \cite{Torino1,Torino2,Torino3}, since these are most 
relevant for comparisons to our results).
These latter authors mapped one aspect of the problem onto an effective non-interacting Anderson localization problem \cite{Torino1}, and thereby found a simple analytical device to obtain (numerically) the Mott insulator-Bose glass phase  very accurately, albeit
within single-site theory. Then, to obtain the Bose glass-to-superfluid transition they analyzed the superfluid stiffness.  They
distinguished the Bose glass phase from that of the superfluid by stating that the condensate fraction, $f_c$, is non-zero
in both of these phases, while the superfluid stiffness is non-zero only in the superfluid region.
However, unlike the enormous lattices that were used by these authors to obtain the Mott insulator-to-Bose glass transition 
lines, when examining the Bose glass and superfluid phases one must solve a set of self consistency equations for
each site on a lattice, and this limits the sizes of the systems that can be studied. Therefore, one is faced with
the difficulties of completing calculations for many lattice sizes, thereby allowing for a finite-size scaling analysis to be completed. Also,
one must average over many different complexions of disorder. Unlike the statements in their paper \cite{Torino1}, when we completed
comparative single-site numerics in the Bose glass and superfluid phases, we found considerable finite size effects, as well as substantial variations with differing
complexions of disorder. As we discuss below, a proposal exists that could explain this behaviour. 

The potential pitfalls of finite-size scaling and averages over complexions of disorder are avoided in  the very elegant and creative theory of Bissbort and Hofstatter, so-called stochastic mean-field theory \cite{SMFT1,SMFT2}. These authors have 
found that for (and, so far,  only for) single-site mean-field theory one may perform exactly (albeit numerically) the averaging over different complexions of disorder in the thermodynamic limit. Unfortunately,
as discussed in detail in Refs. \cite{SMFT2,SMFTthesis}, the stochastic mean-field theory method is one that can only examine the thermodynamic limit. These authors argue, in direct contrast to the work of Refs. \cite{Torino1,Torino2}, that the condensate fraction in the Bose glass phase is identically zero, a result
intimately tied to the existence of rare Lifshitz regions. 
Part of the motivation for our work is a comparison to the finite systems studied experimentally, for example the observed Bose glass phase in three
dimensions \cite{BGexp}, as it is important to be able to understand large but finite systems.  

We have completed large-scale numerical studies of both the single-site approach of Refs. \cite{Torino1,Torino2} as well as the
stochastic mean-field theory of Refs. \cite{SMFT1,SMFT2}, and have found differences between these theories. Therefore, this provided further motivation to go beyond single-site theory to see if the correct physics could be identified. 

There is also a comparison of the single-site mean-field theory with Quantum Monte Carlo studies available in the literature \cite{Zakrzewski,QMC}.
Comparing both position and momentum distributions of particles (including the presence of a harmonic trapping potential),
it has been shown that for a three dimensional Bose-Hubbard model the mean-field approximation results closely resemble the exact solution when the superfluid fraction is non-zero.
It is suggested that the lack of particle-particle correlations in the single-site formulation of the mean-field theory is a deficiency that seems to affect primarily Mott insulator and Bose glass phases.

The idea of extending single-site theories to multiple sites, or clusters, is not, of course, original. In a sense the same idea has been used, \textit{e.g.}, when
the very successful DMFT (dynamical mean-field theory) \cite{dmftRMP} was broadened to cellular or extended DMFT \cite{dmftEXT,dupuis}. That is, 
such work represents an attempt to include more spatial correlations into a theory that ignores the wave-vector
dependence of the self-energy. The same cluster extension has also been adapted to DMFT for disordered systems -- see the paper
of Potthoff  and Balzar \cite{Potthoffdisorder}  in which the Luttinger self-energy functional expansion technique is employed. Also,
a similar idea was applied to the diatomic Bose-Hubbard model, as we outline below. Therefore, the inclusion of such effects is a natural extrapolation from site-decoupled mean-field theories that have been applied in other circumstances before.

So, as discussed, the familiar limitations of mean-field theory apply to site-decoupled mean-field theory, notably the absence of spatial correlations.
For example, for the Bose-Hubbard model one can't calculate pair correlations of the order parameter at differing sites as this corresponds to simply the one-body density matrix.
Our contribution is the extension of site-decoupled mean-field theory for the disordered Bose-Hubbard model to a cluster, or equivalently multi-site mean-field theory.
Our theory treats exactly the spatial correlations on each cluster (each cluster has open boundary conditions), thereby giving us access to spatial correlations over length scales up to the largest distances
spanning a cluster. In one dimension this allows for us to have clusters up to a size of six (eight for homogeneous lattices), whereas for two dimensions even 
though one can examine clusters up to a size of 3$\times$3 for homogeneous clusters \cite{Tom}, for spatially disordered
systems we were required to work with clusters of size 2$\times$2, although one can also work with clusters of 3$\times$2 sites using the technique of basis truncation \cite{Roth}, which is described later in this paper.
Like the site-decoupled single-site mean-field theory discussed
above, in our multi-site work for disordered systems we also allow for a different (self-consistent) order parameter at every site in the lattice.

Our paper is organized as follows. In the next section we present the formalism associated with multi-site mean-field theory applied to the disordered Bose-Hubbard model. In this section we also include the formal details associated with the spatially-averaged variant of multi-site mean-field theory, a natural extension of Ref. \cite{Krutitsky}. In the following section
we present our numerical results of applying this theory to the two-dimensional disordered Bose-Hubbard model, including an extrapolation
of the spatially-averaged variant of single-site mean-field theory of Ref. \cite{Krutitsky} to a multi-site formulation. This is followed in the last section
by a discussion of our results, their relation to previously published data, and further extensions that are possible with this theory.

\section{\label{sec:formalism}Formalism}

The Hamiltonian under study is the disordered Bose-Hubbard model, which is defined (in the grand canonical ensemble)
as
\begin{equation}\label{eq:dBH}
\hat{\mathcal{K}} = \hat{\mathcal{H}} - \mu\hat{\mathcal{N}} = 
 -t \sum_{\langle i,j \rangle} \left( \hat{a}^\dagger_i \hat{a}_j + H.c. \right)
 + \sum_{i} \left( \varepsilon_{i} - \mu\right) \hat{n}_{i}
 + \frac{U}{2}\sum_i \hat{n}_{i}(\hat{n}_{i} - 1)
\end{equation}
where $i,j$ label the sites of the optical lattice, $\hat {a}_{i}$ and $\hat {a}^\dagger_{i}$ annihilate and create a boson on
site $i$, respectively, $\langle i,j \rangle$ denotes that $i$ and $j$ are nearest neighbours, and the number operator for site $i$ is denoted by $\hat{n}_{i}~=~\hat {a}^\dagger_{i}\hat {a}_{i}$. 
The near-neighbour hopping frequency is denoted by $t$, and all hopping beyond nearest neighbours is suppressed. The
on-site (Hubbard) interaction energy is denoted by $U$, and consistent with recent experiments \cite{Ceperley}, interactions
between bosons on different sites are considered to be small and are ignored. The disorder is introduced into the
model considered here via the on-site energies $\varepsilon_{i}$, and we focus on a uniform distribution
of energies from the range of $-W/2$ to $+W/2$.
The chemical potential $\mu$ controls boson number. In what follows below, we restrict our considerations
to zero temperature.

\subsection{\label{subsec:sitedecoupled}Site-decoupled mean-field theory:}

Our work examines the application of a multi-site mean-field theory to this model. To understand the differences with
the familiar and oft-employed site-decoupling mean-field theory (\textit{e.g.}, see Ref.~\cite{Torino1}), note that this latter approximation can be derived
by writing
\begin{equation}
\label{eq:mfdecoupling}
\hat{a}_{i} = \phi_i + \delta \hat {a}_{i}\;\; \text{where~~}\; \phi_{i} \equiv \langle \Psi_0 | \hat{a}_{i} | \Psi_0 \rangle
\end{equation}
with $|\Psi_0\rangle$ representing the ground state, and the (spatially homogeneous) superfluid order parameter being denoted by $\phi$. Then, ignoring terms in the hopping 
Hamiltonian that are second order in $\delta \hat {a}$,  the resulting
model Hamiltonian (denoting site-decoupling by $sd$) for a homogeneous lattice, where all $\varepsilon_{i}=0$ and therefore the order parameter is independent
of lattice site, namely $\phi_{i}=\phi$, is given by
\begin{equation}\label{eq:dBH-sitedecoupling}
\hat{\mathcal{K}}^{sd} = \sum_{i} \hat{\mathcal{K}}^{sd}_{i}
\end{equation}
where
\begin{equation}\label{eq:dBH-sitedecoupling-sitei}
\hat{\mathcal{K}}^{sd}_{i} = \frac{U}{2} \hat{n}_{i}(\hat{n}_{i} - 1) - \mu\hat{n}_{i} - zt \left( \hat{a}_{i} + \hat{a}^\dagger_{i} \right) \phi
\end{equation}
where $z$ is the coordination number for the lattice under consideration. Equation~(\ref{eq:dBH-sitedecoupling}) makes clear that this
formulation leads to a sum over single-site Hamiltonians. Further, this means that the form of the ground-state
wave function is
\begin{equation}\label{eq:Psi-as-product}
| \Psi_0 \rangle = \prod_{i} | \psi_i \rangle
\end{equation}

When one considers a disordered system, the order parameters will, in general, all be different (except in the Mott insulating regions,
for which the order parameter $\phi_{i}$ is zero everywhere). Then one must generalize Eq.~(\ref{eq:dBH-sitedecoupling-sitei}) to be
\begin{equation}\label{eq:dBH-sitedecoupling-sitei-disorder}
\hat{\mathcal{K}}^{sd}_{i} = \frac{U}{2} \hat{n}_{i}(\hat{n}_{i} - 1)
+(\varepsilon_{i}-\mu)\hat{n}_{i} - t \left( \hat{a}_{i} + \hat{a}^\dagger_{i} \right) \sum_{\delta} \phi_{i+\delta}
\end{equation}
where the last sum is over $\delta$ a sum over all sites within one near-neighbour distance to site $i$.
Therefore, one again obtains a Hamiltonian that is site decoupled ($\hat {\mathcal K}^{sd}_{i}$ depends on creation and
annihilation operators only for site $i$), and a ground-state wave function of the form of Eq.~(\ref{eq:Psi-as-product}). However, such a formulation does include spatial correlations in
the sense that for a given complexion of disorder (that is, for a given set of on-site energies $\varepsilon_{i}$), one must solve
the self consistency equation at every site. That is, using Eq.~(\ref{eq:Psi-as-product}) the site-dependent order parameter is found from
\begin{equation}\label{eq:phi-i-SC}
\phi_{i} = \langle \Psi_0 | \hat {a}_{i} | \Psi_0 \rangle = \langle \psi_i | \hat {a}_{i} | \psi_{i} \rangle
\end{equation}
and as is seen from Eq.~(6), the ground state of each $\hat {\mathcal K}^{sd}_{i}$ depends on the order parameters on neighbouring sites.

Last, we note the well-known result \cite{SC} that the minimum value of the grand potential (at $T=0$)
\begin{equation}\label{eq:Omega}
\Omega^{sd} = \langle \Psi_0 | \hat {\mathcal K}^{sd} | \Psi_0 \rangle
\end{equation}
is achieved when the self-consistency conditions given in Eq.~(\ref{eq:phi-i-SC}) are satisfied.
 
\subsection{\label{subsec:msmft}Multi-site mean-field formulation:}

The multi-site version of mean-field theory for the homogeneous Bose-Hubbard model has been stated previously in the 
literature \cite{TorinoMSMFT}, and these authors used this formulation to examine a one-dimensional diatomic lattice.
They obtained data by analyzing one diatomic ``cluster" and examined the stability of so-called loop-hole phases. 
In a subsequent paper these authors stated that the convergence with cluster size (see below) was poor, and therefore
preferred to examine their problems using different approaches. The work of McIntosh, Zaremba and one of the present 
authors \cite{Tom} showed how one obtains improved phase boundaries when one uses a multi-site formulation, in particular
when examining two and three-dimensional lattices. Further, in the analysis of the one-dimensional diatomic lattice the stability conditions
on various ``loop-hole", or ``fractional lobe", phases and mass-density wave states were seen to be changed when more than one diatomic cluster was considered. Below we discuss the extrapolation of these ideas to disordered systems.

The simplest derivation of this formulation can be given in one dimension, and for clarity we detail this formalism for such situations. 
Consider a one-dimensional chain with $N_s$ lattice sites, the sites being labelled by $i = 0, 1, \dots, N_s-1$, with periodic boundary 
conditions. Let the chain be decomposed into
$N_c$ clusters with $L$ sites per cluster, and $N_s = N_c L$. Label the clusters by $I = 0, 1, \dots, N_c-1$, and note
that any given lattice site can now be indexed by $i=\ell+IL$ with $\ell~=~0, 1,\dots, L-1$. Then, write the (exact) grand Hamiltonian as
\begin{equation}\label{eq:dBH-1dclusters}
\hat{\mathcal{K}} = \sum_{I} \hat{\mathcal{K}}^\circ_{I} + \sum_{\langle I,J \rangle} \hat{\mathcal{V}}_{I,J}
\end{equation}
where $\hat{\mathcal{K}}^\circ_{I}$ is given by
\begin{eqnarray}\label{eq:K0-I}
\hat{\mathcal{K}}^\circ_{I} = &\sum_{\ell=0,\dots,L-1}& \left[
\frac{U}{2}\hat{n}_{\ell+IL}(\hat{n}_{\ell+IL} - 1) + 
\left(\varepsilon_{\ell+IL} - \mu\right) \hat{n}_{\ell+IL} \right] \nonumber \\
-t&\sum_{\ell=0,\dots,L-2}& \left( \hat{a}^\dagger_{\ell+IL} \hat{a}_{\ell+1+IL} + \hat{a}^\dagger_{\ell+1+IL} \hat{a}_{\ell+IL} \right)
\end{eqnarray}
Each $\hat{\mathcal{K}}^\circ_{I}$ is the exact grand Hamiltonian for a cluster of length $L$ with open boundary conditions. Therefore,
one should think of Eq.~(\ref{eq:dBH-1dclusters}) as representing a sum of single-cluster Hamiltonians ($\hat{\mathcal{K}}^\circ_{I}$)
that are then coupled via hopping terms between the near-neighbour clusters ($\hat{\mathcal{V}}_{I,J}$). 
One can think of the inter-cluster interactions as being perturbations on the sum over cluster Hamiltonians. 

The idea behind the multi-site mean-field approximation is to apply mean-field decouplings to the inter-cluster hopping terms, $\hat{\mathcal{V}}_{I,J}$. That is, applying Eq.~(\ref{eq:mfdecoupling}) 
only to these terms, and denoting the order parameters at each end of the cluster by $\phi_{i}$, where $i=L-1+IL$ or $i = L+IL$ for
the left and right ends of the cluster,
one finds that the multi-site mean-field approximation (denoted by the superscript $ms$) for the inter-cluster interaction term is given by
\begin{equation}
\hat{\mathcal{V}}^{ms}_{I,I+1} = -t \left[ 
\left( \hat{a}_{L+IL} + \hat{a}^\dagger_{L+IL} \right) \phi_{L-1+IL} +
\left( \hat{a}_{L-1+IL} + \hat{a}^\dagger_{L-1+IL} \right) \phi_{L+IL} 
\right]
\end{equation}
Then, the multi-site mean-field Hamiltonian is seen to be given by
\begin{equation}\label{eq:dBH-msmft}
\hat{\mathcal{K}}^{ms} = \sum_{I} \hat{\mathcal{K}}^\circ_{I} + \sum_{\langle I,J \rangle} \hat{\mathcal{V}}^{ms}_{I,J}
\end{equation}

The main difference between the site-decoupled and multi-site mean-field theories will be made clear below. For now, simply note the 
differing forms of the ground-state wave functions. For the site-decoupled formulation one obtains the form shown in 
Eq.~(\ref{eq:Psi-as-product}), whereas for the multi-site formulation one has
\begin{equation}\label{eq:Psi-as-cluster-product}
| \Psi_0 \rangle = \prod_{I} | \psi_I \rangle \end{equation}
where $ | \psi_I \rangle$ is an energy eigenstate of the cluster Hamiltonian $\hat{\mathcal{K}}^\circ_{I}$. That is, the eigenstates
of $\hat{\mathcal{K}}^\circ_{I}$
include the spatial correlations present in a fully interacting cluster of length $L$ with open boundary conditions, and specifically one
sees that these eigenstates depend on the hopping parameter, $t$.  If instead one treats the hopping
as a perturbation in the site-decoupled formulation, the unperturbed eigenstates do not depend on the hopping frequency. This begs the question,
does the inclusion of some effects of hopping into the zeroth-order eigenvectors lead to more accurate estimates
of the phase boundaries and quantum phases? This was, in part, our motivation for this study.

\begin{figure}[ht!]
\includegraphics[scale=0.35]{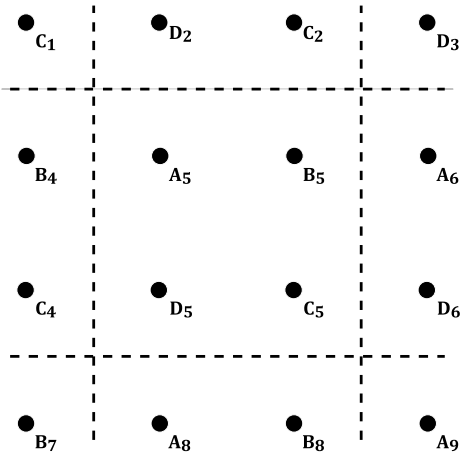}
\caption{ \label{fig:cluster}
A portion of a two-dimensional square lattice that is partitioned into 2$\times$2 clusters -- only the central cluster is shown
in its entirety. The dashed lines show the division of the lattice into such clusters. In each cluster there are four sites,
labelled by \textbf{A,B,C,D}, and each cluster is denoted by a subscript, which runs from 1 to 9.}
\end{figure}

The geometry associated with the generalization of this idea to higher dimensions, and in particular to the focus of this paper, namely
the two-dimensional square lattice, is straightforward -- the site and cluster labels are now vectors, and an example
partitioning of the lattice is shown in Fig.~\ref{fig:cluster}. The only technical complication
is that every site on the ``edge" of a cluster has its own order parameter, and for a disordered system all such order parameters
are, in general, different.

An important difference is seen immediately -- the zeroth-order cluster Hamiltonian for a site-decoupled theory does not depend
on the hopping frequency, $t$, and does not include any spatial correlations between the (disordered) on-site energies.  

The form of the cluster states that are eigenvectors of $\hat{\mathcal{K}}^\circ_{I}$ can be written as
\begin{equation} \label{eq:basisset}
| \psi_I \rangle = \sum_{n_0=0}^{n_{\max}} \sum_{n_1=0}^{n_{\max}} \ldots \sum_{n_{L-1}=0}^{n_{\max}} A_{n_0,n_1,\ldots,n_{L-1}}
(\hat{a}^\dagger_{n_0})^{n_0}(\hat{a}^\dagger_{n_1})^{n_1}\ldots(\hat{a}^\dagger_{n_{L-1}})^{n_{L-1}} |0\rangle
\end{equation} 
and numerically one must choose some finite $n_{\max}$ that leads to fully converged results.
We have found that for the region of the phase diagram associated with the first Mott lobe, namely $0 \leq \tilde \mu \leq 1$,
$n_{\max}=3$ is sufficient when $\tilde t \lesssim 0.08$ -- all of our results shown in this paper are within this range.

The dimensionality of the Hilbert space grows very rapidly with increasing cluster size and so does the computational time.
Therefore, in order to go beyond clusters of 2$\times$2 sites, one needs to reduce (truncate) the number of basis states even further.
The challenge is to decide which states are important and should be retained and which can be discarded.
Of course, one still wants the results to be converged. Following the work of Schmitt, \emph{et al.} \cite{Roth} we have adopted 
a truncation scheme which allows for a physically motivated and controlled approach that discards those states which contribute the least 
to the particular cluster's ground state. As a simple measure for the importance of individual number states one can use the 
expectation value of the Hamiltonian. The key idea is to discard only those basis states for which the expectation value 
of the Hamiltonian is greater than some value.
Those states have a considerably smaller contribution to the final ground state of a particular cluster, and
this procedure leaves us with a given percentage of the initial basis set.
We found that for clusters of 3$\times$2 sites it is enough to retain only 30\% of the ``full" basis set, ``full" corresponding to $n_{\max}=3$ of Eq.~(\ref{eq:basisset}) (as discussed in the previous paragraph).
Having in mind that the truncation scheme works well \cite{Roth} in the regions where the ground state is composed of few dominant basis states,
one has to remember that the results for the superfluid phase will be less reliable.
(One needs the full real space basis set to accurately describe the superfluid phase, since it is a phenomenon observed in the momentum space.)

\subsection{Quantum phases:}

We will work in the strong-coupling limit in that we assume that $U$ is the largest energy scale, and instead of working
with $\hat{\mathcal{K}}$ and its mean-field variants we work with the dimensionless
\begin{equation}
\hat{\mathcal{K}} \rightarrow \hat{\tilde{\mathcal{K}}} = \frac {\hat{\mathcal{K}}} {U}
\end{equation}
and also express the various material parameters as scaled energies, namely
\begin{equation} \label{eq:scaling}
t \rightarrow \tilde t = \frac {t}{U} \;\;\;\;\;
\mu \rightarrow \tilde \mu = \frac {\mu}{U} \;\;\;\;\;
\varepsilon_{i} \rightarrow \tilde \varepsilon_{i} = \frac {\varepsilon_{i}}{U}
\end{equation}
Letting the (scaled) energy $\tilde W$ characterize the strength of the disorder, the phase diagram for this
model can be obtained in $(\tilde t, \tilde \mu, \tilde W)$ space. To begin, we present a brief discussion of
the phases that appear in such phase diagrams.

For a homogeneous lattice with only near-neighbour hopping and on-site repulsion, one finds that Mott insulating phases
are found to correspond to an integral number of bosons per site. The superfluid phase is then found in the $(\tilde t, \tilde \mu)$
plane to correspond to $\phi \not= 0$. For non-homogeneous lattices the situation is more complicated. 
Specifically, when $\phi = 0$ one finds that the Mott insulator phase is an incompressible phase that 
corresponds to integral densities of bosons per lattice site. For ordered non-homogeneous lattices, for example for a diatomic 
lattice in one dimension, one finds a variety of density-wave phases \cite{Rousseau,Tom}, (so that the average
density per site is integral, but the local bosonic densities need not be integral) that are also incompressible and
therefore Mott insulators. In addition to such behaviour, even for binary disorder one finds Mott
insulating phases but now of a different nature \cite{Mering};  these incompressible phases do not even have an average
boson density per site that is integral. Our analysis focuses on non-homogeneous
lattices with a disorder that is not binary (from now on in our discussion it is implicit that we are not
addressing the situation of binary disorder), and the transition to non-Mott insulating phases is qualitatively different.

Formally, one notes that the exact grand Hamiltonian of 
Eq.~(\ref{eq:dBH}) commutes with the total number operator, and therefore in the absence of degeneracies or 
symmetry-breaking terms being added to Eq.~(\ref{eq:dBH}), the eigenstates are also number eigenstates.
For systems with spatially disordered on-site energies, $\tilde \varepsilon_{i}$, this is also true of both the site-decoupled 
(Eq.~(\ref{eq:dBH-sitedecoupling})) and multi-site (Eq.~(\ref{eq:dBH-msmft})) mean-field Hamiltonians when the order 
parameters $\phi_{i}$ are zero on \textit{all} sites.  However, when the mean-field terms are included they act
as a source of symmetry-breaking interactions, and now the ground state is found to not correspond to number
eigenstates (either for a single site or for a multi-site ground states -- see Eqs.~(\ref{eq:Psi-as-product},\ref{eq:Psi-as-cluster-product})).
Therefore, the resulting phases are compressible. Also, as seen from Eq.~(8), in such a situation one obtains
self-consistent solutions for which $\phi_i \not= 0$.

For a homogeneous lattice the compressible phase is known to be a superfluid. However, for spatially disordered systems
it is known that between the Mott insulating and superfluid phases a new quantum state arises, a phase called
the Bose glass phase. Several important results concerning this transition were summarized in the introduction to
this paper. Here we present the methodology that we have employed to determine the phase boundary separating
the Mott insulating and Bose glass phases in $(\tilde t, \tilde \mu, \tilde W)$ space. 

To identify the phase boundary of the Mott insulating and superfluid phases for the homogeneous system one may use
a variety of methods. For example, one may use perturbation theory \cite{Vanoosten} on a Landau-type expansion of the 
grand potential $\Omega$ in the order parameter $\phi$. Simply, when the coefficient of the $\phi^2$ term is found to change
from being positive to negative (for say constant $\tilde \mu$ with increasing $\tilde t$), the order parameter becoming
non-zero signifies a continuous phase transition from the Mott insulator to superfluid phases. This is the method that
McIntosh, Zaremba and one of us have used \cite{Tom} in analyzing various ordered systems.

A variant of this approach has been used for disordered systems by treating the site-decoupled mean-field theory
with a spatial average over disorder \cite{Krutitsky}. To be concrete, \textit{if} one approximates the order
parameter to be spatially homogeneous for the disordered system, namely $\phi_{i} \rightarrow \overline \phi$, 
and ignores the correlations between neighbouring sites, so that $\phi_{i} \phi_j \rightarrow \left( \overline \phi \right)^2$, 
then using the same approach as for a homogeneous system (for a given distribution of on-site energies) in a site-decoupled
mean-field approximation one may derive \cite{Krutitsky} (in fact, analytically) the loss of stability of the Mott insulating phase with
increasing disorder. Below we explain how this approach can be extended to our multi-site mean-field formulation; such results
serve as the most simplistic approach in using our formalism to determine the locations of the phase boundaries.

The approach that utilizes the full spatial structure of the on-site energies and order parameters that we have used
can be explained as follows. Again for clarity, consider a one-dimensional lattice with $L$ sites per cluster. For a given 
complexion of disorder any lattice will have different values of the order parameter at every site (outside of the $\phi = 0$ 
Mott insulator region). Then the phase boundary in such an approach can be derived by consideration of a stability matrix.  
That is, beginning with 
\begin{eqnarray}\label{eq:Ktildems}
\hat{\tilde{\mathcal{K}}}^{ms} &=& \sum_{I} \hat{\tilde{\mathcal{K}}}^\circ_{I} + \sum_{\langle I,J \rangle} \hat{\mathcal{V}}^{ms}_{I,J}\\
&=& \sum_{I} \hat{\tilde{\mathcal{K}}}^\circ_{I} + {\tilde t} ~ \hat{\mathfrak v}\nonumber
\end{eqnarray}
where (for simplicity with the notation, restricting our attention to one-dimensional chains)
\begin{equation}\label{eq:v-msmft}
 \hat{\mathfrak v} = - \sum_{I} \left[ 
\left( \hat{a}_{L+IL} + \hat{a}^\dagger_{L+IL} \right) \phi_{L-1+IL} +
\left( \hat{a}_{L-1+IL} + \hat{a}^\dagger_{L-1+IL} \right) \phi_{L+IL}
\right]
\end{equation}
Then, representing the decoupled (that is, $\hat{\mathfrak v} = 0$) eigenstates via
\begin{equation}\label{eq:K0eigs}
\sum_{I} \hat{\tilde{\mathcal{K}}}^\circ_{I} | \Psi^\circ_\alpha \rangle = E^\circ_\alpha | \Psi^\circ_\alpha \rangle
\end{equation}
with all eigenstates of Eq.~(\ref{eq:K0eigs}) being labelled by the index $\alpha$, and 
$E^\circ_0$,~$| \Psi^\circ_0 \rangle$ being the ground-state energy and eigenvector of the decoupled system  respectively, one determines the ground state of $\hat{\tilde{\mathcal{K}}}^{ms}$ to lowest order in the perturbation $\hat{\mathfrak v}$, namely
\begin{equation}
| \Psi_0\rangle = | \Psi^\circ_0 \rangle + | \Psi^{(1)}_0 \rangle 
\end{equation}
where
\begin{equation}
| \Psi^{(1)}_0 \rangle = {\tilde t}~\sum_{\alpha\not=0} 
\frac{\langle \Psi^\circ_\alpha |  \hat{\mathfrak v} |  \Psi^\circ_0 \rangle}{E^\circ_0 - E^\circ_\alpha}
| \Psi^\circ_\alpha \rangle
\end{equation}
The self-consistency equation, to lowest order in $\hat{\mathfrak v}$, can then be written as
\begin{eqnarray} \label{eq:stabilitymatrix}
\phi_{i} = \langle \Psi_0 | {\hat a}_{i} | \Psi_0 \rangle &\approx&
\left(\langle \Psi^\circ_0 | {\hat a}_{i} | \Psi^{(1)}_0 \rangle +
\langle \Psi^{(1)}_0 | {\hat a}_{i} | \Psi^\circ_0 \rangle \right) \nonumber\\
&=& \sum_{j} {\mathbb C}_{ij} \phi_{j}
\end{eqnarray} 
since one notes, from Eq.~(\ref{eq:v-msmft}), that $\hat{\mathfrak v}$ varies linearly with $\phi_{i}$. This quantity, ${\mathbb C}$, is what
we refer to as the stability matrix. That is, when the magnitude of any eigenvalue of ${\mathbb C}$ exceeds one, the Mott-insulating phase, 
corresponding to $\phi_{i} = 0~\forall {i}$, is found to be unstable. 

For the homogeneous lattice one finds that the Mott insulating phase is unstable to the formation of a superfluid. However, for the
disordered Bose-Hubbard model one finds an intervening Bose glass phase. As is well known \cite{Penrose,Fisher89}, the appearance of
the condensate in an interacting system may be identified by a non-zero condensate fraction, denoted by $f_c$, which may be evaluated based
on the largest eigenvalue of the one-body density matrix in the thermodynamic limit.  This latter quantity is defined as follows. First, let $N_b$ denote the total number of bosons, namely
\begin{equation}
N_b = \sum_{i} n_{i}  = \sum_{i} \langle \Psi_0 | {\hat n}_{i} | \Psi_0 \rangle
\end{equation}
Then, the one-body density matrix may be defined by
\begin{equation}
\rho^{\rm I}_{ij} = \frac {\langle \Psi_0 | {\hat a}^\dagger_{i} {\hat a}_{j}| \Psi_0 \rangle}{N_b}
\end{equation}
Now consider the multi-site mean-field theory for a one-dimensional crystal, and, to be concrete, consider a cluster size of $L=2$.
Noting that when the order parameters $\phi_{i}$ are non-zero the ground state wave function is still of the same form
as in Eq.~(\ref{eq:Psi-as-cluster-product}), it is then simple to show that
\begin{equation}
\rho^{\rm I} = \frac{1}{N_b}
\begin{pmatrix}
n_0 & \langle {\hat a}^\dagger_0 {\hat a}_1\rangle & \phi_0 \phi_2 & \phi_0 \phi_3 & \dots & \phi_0 \phi_{N_s-2} & \phi_0 \phi_{N_s-1} \\
\langle {\hat a}^\dagger_1 {\hat a}_0\rangle & n_1 & \phi_1 \phi_2 & \phi_1 \phi_3 & \dots & \phi_1 \phi_{N_s-2} & \phi_1 \phi_{N_s-1} \\
\phi_2 \phi_0 & \phi_2 \phi_1 & n_2 & \langle {\hat a}^\dagger_2 {\hat a}_3\rangle & \dots & \phi_2 \phi_{N_s-2} & \phi_2 \phi_{N_s-1} \\
\phi_3 \phi_0 & \phi_3 \phi_1 & \langle {\hat a}^\dagger_3 {\hat a}_2\rangle & n_3 & \dots & \phi_3 \phi_{N_s-2} & \phi_3 \phi_{N_s-1} \\
\vdots & \vdots & \vdots & \vdots & \ddots & \vdots & \vdots \\
\phi_{N_s-2} \phi_0 & \phi_{N_s-2} \phi_1 & \phi_{N_s-2} \phi_2 & \phi_{N_s-2} \phi_3 & \dots & n_{N_s-2} & \langle {\hat a}^\dagger_{N_s-2} {\hat a}_{N_s-1}\rangle \\
\phi_{N_s-1} \phi_0 & \phi_{N_s-1} \phi_1 & \phi_{N_s-1} \phi_2 & \phi_{N_s-1} \phi_3 & \dots & \langle {\hat a}^\dagger_{N_s-1} {\hat a}_{N_s-2}\rangle & n_{N_s-1}
\end{pmatrix}
\end{equation}
Note that for the homogeneous system all order parameters are equal, and one has that $N_b = n N_s$. 
Therefore the largest eigenvalue of this matrix would then be given by
\begin{equation} 
\label{eq:trouble}
\lambda_{\max} = \frac{1}{N_s} + \frac{\langle {\hat a}^\dagger_0 {\hat a}_1\rangle}{nN_s} + \frac{(N_s - 2)\phi^2}{nN_s}
\end{equation}
and thus
\begin{equation} 
\label{eq:fc}
f_c \equiv \lim_{N_s\rightarrow \infty} \lambda_{\max} = \frac{\phi^2}{n}
\end{equation}
as expected. 
(Identical results are obtained for any homogeneous system treated with multi-site mean-field theory, with any cluster size.)
For the disordered system, the largest eigenvalue can be extracted numerically, and finite-size scaling should allow for one to obtain $f_c$.
The scaling form follows from Eq.~(\ref{eq:trouble}), and below
and in our figures we denote this finite-size corrected condensate fraction to be denoted by $\tilde f_c$.

The quantum phase that appears at the Mott-insulating phase boundary is known to be the Bose glass phase \cite{Fisher89}. The condensate 
fraction in this phase, at least for finite systems, is non-zero. However, unlike the homogeneous system, for a disordered system the superfluid stiffness, 
 $\rho_s$, may be zero while the condensate fraction, $f_c$, is non-zero. In the thermodynamic limit this is purported to be associated with islands of condensate that are not phase coherent with one another \cite{SMFT2}.  
Below we show how the complete phase diagram can be (approximately) determined from the condensate fraction alone.
\section{Results for the 2d square lattice with box disorder}

We now turn to the evaluation of the phase boundaries and an examination of the properties of the quantum phases for a two-dimensional square lattice
having spinless bosons described by the disordered Bose-Hubbard model, given in Eq.~(\ref{eq:dBH}),
and present our numerical results for both the site-decoupled single-site
mean-field theory and for our multi-site formulation. 
We have solved this problem for a variety of lattice sizes such that the number of lattice sites times the number of complexions
of disorder is 40,000, and the lattice sizes range up to 200$\times$200 sites. We discuss
the ``convergence" that we obtain with these choices below.
In our multi-site work we use mainly clusters of size 2$\times$2, since
the dimensionality of the Hilbert space of a cluster grows very rapidly with increasing cluster size.
To reduce the number of basis states for homogeneous lattices one can use group theory to find equivalent lattice sites, and hence equivalent basis states \cite{Tom}.
However, in the presence of disorder the C$_4$ point group symmetry is broken, so no such formal device is available.
Therefore, to go beyond clusters of 2$\times$2 sites we have used a basis truncation scheme that was introduced elsewhere \cite{Roth}, and which was described earlier in the previous section of the present work. Through this approach we have been able to extend our study to 
examine the results that follow from the use of clusters of 3$\times$2 sites.
Our results are limited to the region of the phase diagram associated with the first Mott lobe, namely $0 \leq \tilde \mu \leq 1$,
when determining the phase boundaries (see Eq.~(\ref{eq:stabilitymatrix})) and self-consistent solutions for the order parameter at each site of a given lattice.
Also, we present results for the case of a uniform (or box) distribution of on-site energies with $\tilde W = 0.5$ (see Eq.~(\ref{eq:scaling})).
More details on the state labelling (and related numerical issues) can be found in \cite{Zhang}.
 
\begin{figure}[ht!]
\includegraphics{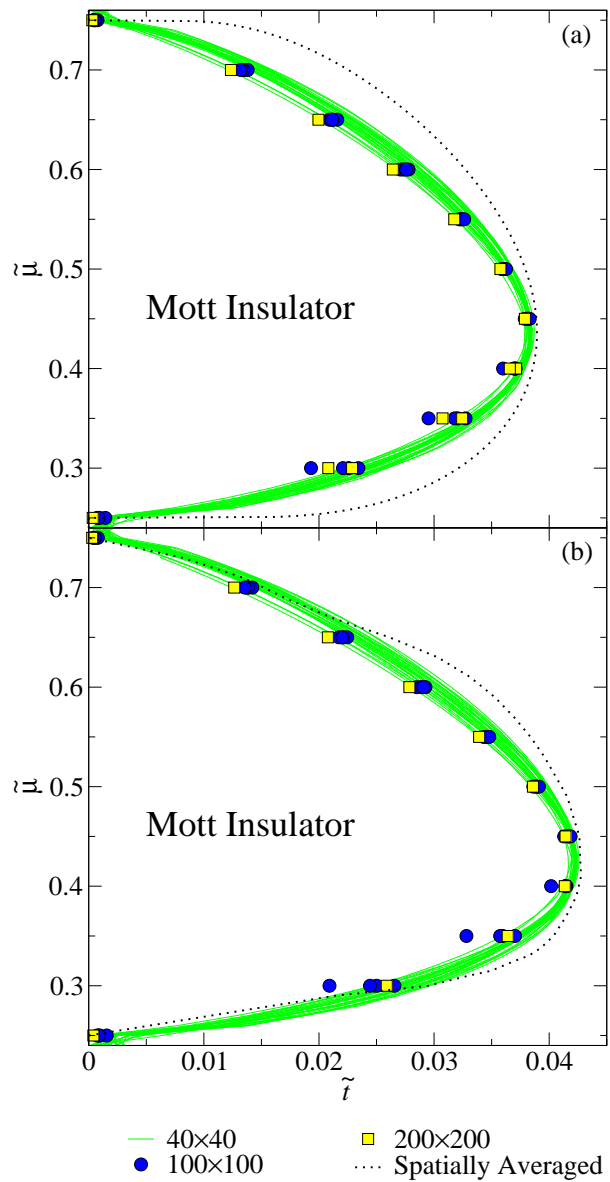}
\caption{  \label{fig:pb}
Mott insulator/Bose glass phase boundaries of a two-dimensional (a) site-decoupled single-site mean-field theory (SSMFT)
and (b) multi-site mean-field theory (MSMFT), where the clusters are 2$\times$2 sites, for a box distribution
of disorder of strength $\tilde W = 0.5$. 
The results shown are for lattice sizes of 40$\times$40, 100$\times$100 and 200$\times$200, with 25, 4 and 1 complexion of disorder, respectively,
and are represented using the line types/point styles given in the legend. For example, we are superimposing 25 stability
curves for the 40$\times$40 lattices.
The spatially-averaged phase boundary is found using the approach of \protect\cite{Krutitsky} for single-site mean-field theory, and as described in the main body of the paper for 2$\times$2 clusters.}
\end{figure}

The Mott insulator-to-Bose glass phase boundary obtained within the chosen range of $\tilde \mu$ is given in Fig.~\ref{fig:pb}.
For the single site mean-field theory (see Fig.~\ref{fig:pb}a) we show three different lattice sizes --
25 complexions of disorder for a 40$\times$40 lattice, 4 complexions of a 100$\times$100 lattice and a single complexion of a 200$\times$200 lattice.
Additionally, for the  200$\times$200  system, we show the phase boundary of a different realization of disorder, but for only three values of the chemical potential;
the chemical potentials used are those for which we saw the 100$\times$100 lattice to have the biggest variations in the phase boundary results.
It is clear from this figure that not only does one obtain a range of phase boundaries at smaller lattice sizes, but also for much larger lattices.
In contrast to the claim of Ref.~\cite{Torino1}, 
who studied many lattices and complexions of disorder in one dimension and found no dependence on either of these (and claimed
that a chain of length 100 did \emph{not} suffer from finite-size effects), and two dimensions in other published work \cite{Torino3}, in our own
numerical study of site-decoupled single-site mean-field theory we find a strong dependence
of these phase boundaries on lattice size and complexions of disorder. 
(However, for the \textit{condensate fraction} for large systems our single-site theory numerics are consistent with their statements  -- see below.)

We show the results for our multi-site formulation in Fig.~\ref{fig:pb}b, with the same number of complexions of disorder for each of the lattice sizes studied for single-site theory.
For all lattice sizes and all complexions of disorder we find that the
area of the phase diagram corresponding to the Mott insulator phase is greater when calculated using our multi-site theory
than that predicted by single-site theory, a result consistent with studies on homogeneous lattices \cite{Tom}.

If one examines Fig.~\ref{fig:pb} at some fixed chemical potential but with an increasing hopping frequency, in particular for chemical potentials in the lower half 
of the Mott lobe (namely around $0.3 \leq \tilde \mu \leq 0.45$), one finds a considerable variation of the location of the phase transition even for our
largest lattices. An identical spread of the phase boundaries is found for single-site theory, even for lattices with 200$\times$200 sites.
This result, that one does not find a convergence to a single phase transition \cite{Sornette}, is not consistent the self averaging that one might expect for large lattices.
Therefore, for most experimental systems (read: finite with less than 40,000 lattice sites), for a given complexion of disorder one can expect some variation in the stability of
the Mott insulating and Bose glass phases, depending on both the extent of the harmonic trap (thus dictating the (approximate) effective size of the
lattice), and on the particular complexion of disorder that is produced in a given experiment. An interesting aspect of this observation is that
it is seen to be consistent with a recent renormalization group analysis of this transition \cite{phillips} -- for the disorder fixed points, they predict the absence of self averaging in this model.

In addition to the numerical results of Fig.~\ref{fig:pb}, one may extend the spatially averaged disorder theory of Krutitsky, \emph{et al.} \cite{Krutitsky}, to the multi-site 
formulation, as we now explain. Referring to the above-outlined perturbation theory, if one wishes to determine the variation of the energy (more
specifically, the grand potential) when the effects of the perturbation are included (see Eq.~(\ref{eq:Ktildems})), as in Refs.~\cite{Vanoosten,Krutitsky}, one obtains a Landau-type
thermodynamic potential for each cluster of the form
\begin{equation} \label{eq:omegai}
\Omega_I = \Omega_I^\circ + \sum_{i,j \in I} A_{i,j} \phi_i\phi_j + \textit{higher-order terms}
\end{equation}
The expansion coefficients depend on the various on-site energies, $\tilde \varepsilon_i$ of Eq.~(\ref{eq:scaling}), and paralleling the approach of Ref.~\cite{Krutitsky}
we approximate that for all $i,j$
\begin{equation}\label{eq:phiiphij}
\overline{\phi_i\phi_j} \rightarrow \left( \overline{\phi} \right)^2
\end{equation}
where the spatially averaged superfluid order parameter is used to replace all local order parameters appearing in Eq.~(\ref{eq:omegai}). Therefore,
the Mott insulator-to-Bose glass transition is found to correspond to the crossing of zero (say with increasing $\tilde t$ at some fixed $\tilde\mu$) of the second-order coefficient.
(Of course, Eq.~(\ref{eq:phiiphij}) is satisfied exactly in the Mott insulator.) For a homogeneous lattice this formalism is presented elsewhere
\cite{Tom}, and for a disordered lattice using  2$\times$2 clusters the extrapolation of \cite{Krutitsky} to our
multi-site formulation
requires that one average over all possible on-site energies on each of the four sites, something that can only be completed numerically. The single-site result of \cite{Krutitsky} is plotted in Fig.~\ref{fig:pb}a, and for the multi-site
theory the resulting phase boundary is shown as a dotted line in Fig.~\ref{fig:pb}b. Notably, it is seen to be much closer to the numerical results described above
for large lattices than is obtained with site-decoupled single-site mean-field theory. That is, by including the spatial correlations within each  
multi-site cluster one obtains a reasonably accurate representation of the MSMFT phase boundary.
(Unlike the single-site theory \cite{Krutitsky}, we are not able to complete the required integrals analytically
and therefore can not write down a simple analytical expression for the phase boundaries of a spatially averaged multi-site theory.) Clearly, with
increased computing power this idea can be extended to larger clusters, thereby providing one with a much better disorder-averaged phase boundary, if the  ${\hat {\tilde {\mathcal K}}}^\circ_I$ eigenstates can be calculated.

\begin{figure}[ht!]
\includegraphics[scale=1.1]{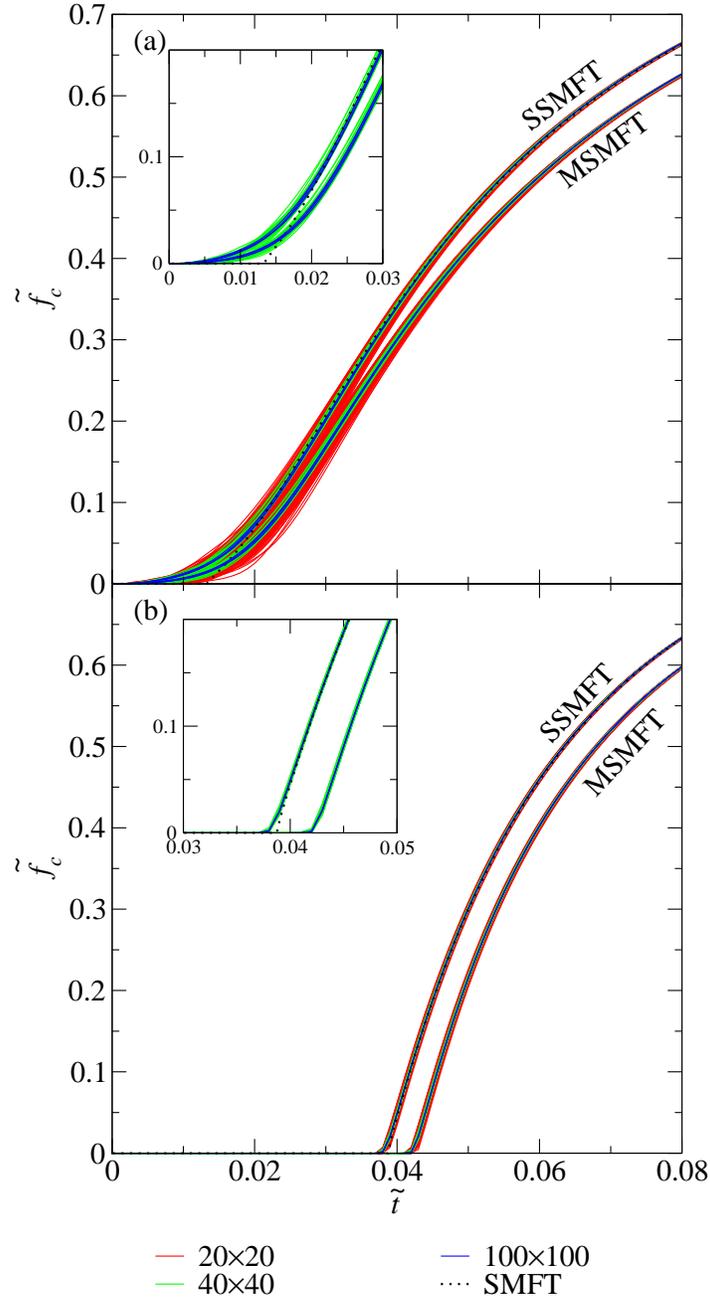}
\caption{  \label{fig:fc} 
The variation of the condensate fraction, $\tilde f_c$, with hopping frequency, $\tilde t$,
for $\tilde W = 0.5$ and (a) $\tilde \mu = 0.1$ and (b) $\tilde \mu = 0.4$ -- the appropriate finite-size corrections for
the condensate fraction are included.
For single-site mean-field theory (SSMFT) the results for lattices of, respectively, 20$\times$20, 40$\times$40, and 100$\times$100 sites with 100, 25 and 4 complexions of disorder are shown.
The legend shows linetypes/colours to which the particular lattice size results correspond.
One sees that (consistent with Ref.~\cite{Torino1}) there is virtually no change in $f_c$ for sufficiently large lattices --
the 40$\times$40 curves are nearly coincident, and the 100$\times$100 curve lies on top of them --
providing that the condensate fraction is not close to being zero.
Note that our results from stochastic mean-field theory (SMFT), denoted by the dotted line, 
agree with the large lattice results for the single-site theory, but only for the condensate fraction $\tilde f_c \gtrsim 0.1$ (see inset).
In addition, we show the results obtained from our multi-site mean-field theory (MSMFT) for clusters of 2$\times$2 sites.
The lattice sizes and numbers of complexions of disorder are the same as studied for single-site theory.
}
\end{figure}

The variation of the condensate fraction with hopping frequency, correcting for finite-size effects, as described at the end of the previous section, is shown in Fig.~\ref{fig:fc} for two values of the chemical potential;
one below the first Mott lobe (which for $\tilde W=0.5$ is found to be bounded by $0.25 \leq \tilde \mu \leq 0.75$), namely $\tilde \mu=0.1$, and one inside, that is $\tilde \mu=0.4$.
Again, both single- and multi-site mean-field theory results are shown 
for a variety of lattice sizes for which the number of lattice sites times the number of complexions
of disorder is 40,000.
We find that in both single- and multi-site formulations of the mean-field theory the condensate fraction seems to converge to a fixed value independent of the complexion of
disorder when large enough lattices are used, but \emph{only} when this fraction
is away from zero; this behaviour is similar to what was found in the numerics of Ref.~\cite{Torino1}.
Also, if one examines these curves closely in the vicinity of the Mott insulator-to-Bose glass phase transition, when
the condensate fraction is very close to zero (for $\tilde\mu=0.4$),
one finds that the condensate fraction becomes nonzero precisely at the values of the hopping frequency
found from the instability theory (see Fig.~\ref{fig:pb}). That is, when $\tilde f_c$ first becomes non-zero for a given lattice size varies with complexion of disorder, and does not converge to a fixed value.
Below we refer to this hopping value as $\tilde t_{c1}$. (For completeness,
note that $\tilde t_{c1}=0$ for $\tilde \mu =0.1$.) As is clear from these plots, we find that the implementation of a multi-site theory 
results in qualitatively similar changes to those of the phase boundaries that are found from the single-site formulation. That is, the onset of $f_c$
(say $f_c$ greater than $\sim 0.01$) is shifted to higher values of $\tilde t$, similar to the shift of MI/BG phase boundary discussed above
-- see Fig.~\ref{fig:pb}.

\begin{figure}[ht!]
\includegraphics{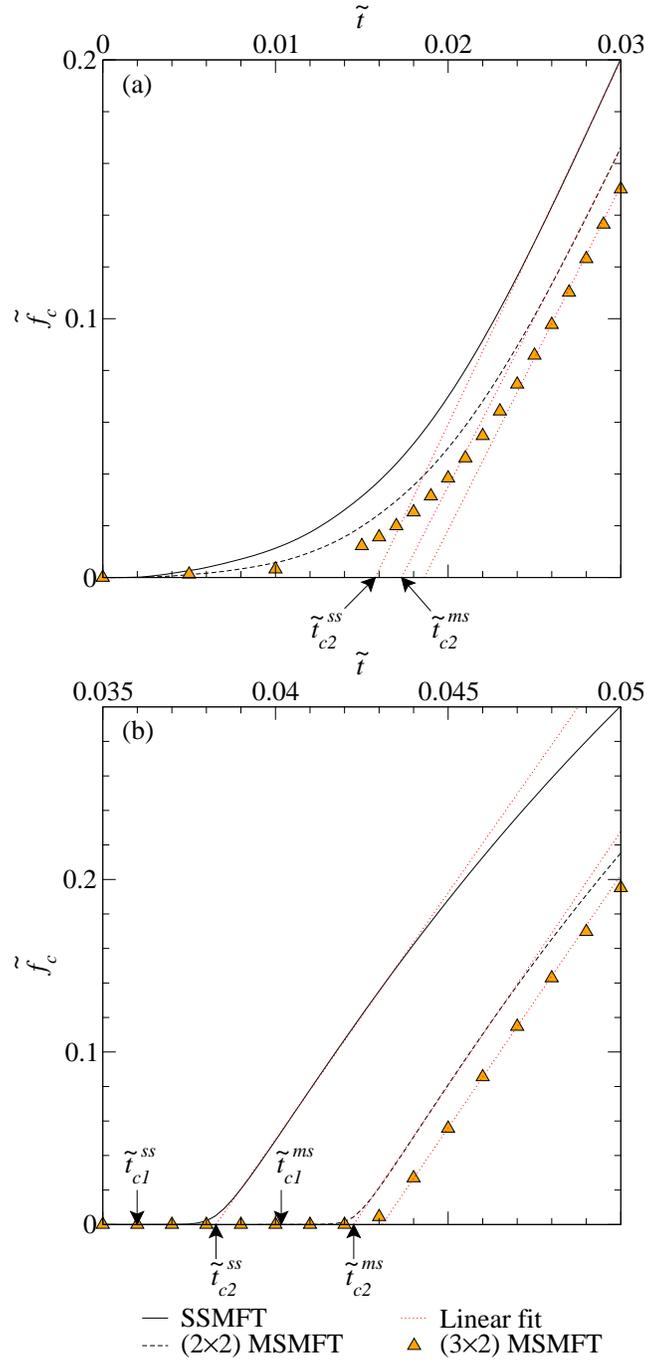}
\caption{ \label{fig:fcsingle}
The variation of the condensate fraction, $\tilde f_c$, for both SSMFT and MSMFT, 
with hopping frequency, $\tilde t$,
for $\tilde W = 0.5$ and (a) $\tilde \mu = 0.1$ and (b) $\tilde \mu = 0.4$ for one complexion of disorder.
Both the SSMFT and (2$\times$2) MSMFT results have been obtained on a 100$\times$100 lattice for the same complexion of disorder.
The results for (3$\times$2) MSMFT have been obtained on a 30$\times$30 sites lattice with 30\% basis truncation (see text for more details).
The linear fit of the data provides an estimate of the Bose glass-to-superfluid phase transition,
$\tilde t_{c2}^{ss}$ and $\tilde t_{c2}^{ms}$, respectively, for SSMFT and MSMFT.
The Mott insulator-to-Bose glass transition at $\tilde t_{c1}^{ss}$ and $\tilde t_{c1}^{ms}$ are also shown for $\tilde \mu = 0.4$; 
for $\tilde \mu = 0.1$ these $\tilde t_{c1}$ quantities are both zero.
}
\end{figure}

The above results have led us to identifying a simplistic approximation that shows how one can estimate not only when the Mott insulator 
loses stability (as discussed above, this corresponds to $\tilde t_{c1}$), but how one can also obtain an estimate  of the phase boundary separating the Bose glass and superfluid phases, considering only the variation of the condensate fraction. 
That is, at larger hopping frequencies the disorder is  found to not suppress phase coherence throughout the lattice (see below),  and one obtains a superfluid. We shall identify the latter transition by $\tilde t_{c2}$. 

To make clear how our approach works, in Fig.~\ref{fig:fcsingle} we show the variation of this quantity for values of 
the chemical potential for which the Mott insulator phase does not exist ($\tilde t_{c1}=0$), \textit{viz.} for $\tilde \mu = 0.1$,
and also for which it is stable only below some critical value of $\tilde t_{c1}>0$, \textit{viz.} for $\tilde \mu = 0.4$.
One can estimate the critical values ${\tilde t}_{c2}^{ss}$ and ${\tilde t}_{c2}^{ms}$, 
at which the system transforms from being in the Bose glass phase to that of a superfluid, as follows.
Recall from above that one may expand the grand potential in terms of a spatially
homogeneous order parameter, $\overline \phi$, phenomenologically as a Landau-type expansion, namely
\begin{equation}
\Omega = \Omega^\circ + \frac{|A|}{2} \left( \tilde t_{c} - \tilde t \right)^\frac12 \left(\overline\phi \right)^2 + \frac{B}{4} \left( \overline\phi \right)^4
\end{equation}
From this it follows that when the order parameter first becomes non-zero, one has that
\begin{equation}
\overline \phi \sim \left( \tilde t_{c} - \tilde t \right)^{\frac12} 
\end{equation}
a common mean-field result.
Then noting that for a homogeneous system the superfluid fraction, $f_s$, is equal to the condensate fraction, our previous result of $\tilde f_c \sim \phi^2$ shows
that one then obtains that the superfluid fraction first develops linearly in the hopping frequency. 
From Fig.~\ref{fig:fcsingle} one sees that when the condensate fraction
first becomes sizable it grows linearly with $\tilde t$, and therefore 
one can find a simplistic approximation for ${\tilde t}_{c2}$ by extrapolating back from the linear region.
The results of this procedure are shown in the figure and are denoted with (red) dotted lines. 
The results shown refer to single-site mean-field theory, $t_{c2}^{ss}$, and multi-site mean-field theory with clusters of 2$\times$2 sites, denoted
by $t_{c2}^{ms}$. 

\begin{figure}[ht!]
\includegraphics{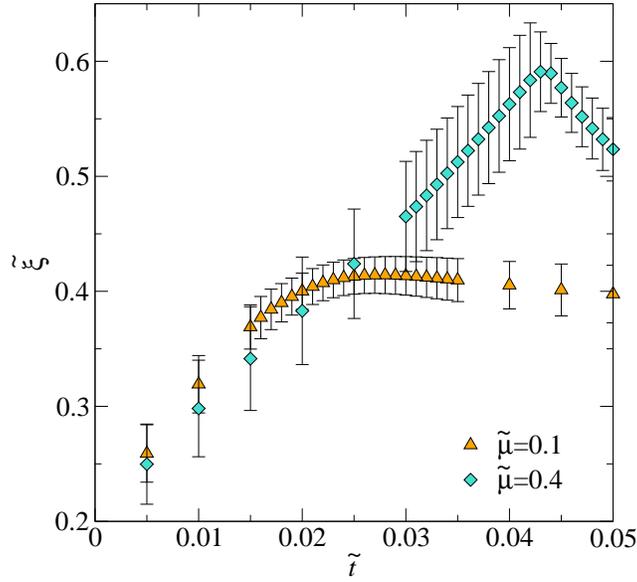}
\caption{ \label{fig:xi}
The variation of the correlation length $\tilde \xi$ (from Eq.~(\ref{eq:correlation})) with hopping frequency $\tilde t$,
for one particular complexion of box disorder of strength $\tilde W = 0.5$, evaluated using our multi-site mean-field theory
with clusters of 3$\times$2 sites.
The results have been obtained on a 30$\times$30 sites lattice with 30\% basis truncation (see text for more details).
The error bars show asymptotic standard error of the correlation length fit parameter. Note that the peak of
the correlation length for $\tilde \mu = 0.4$ agrees with the location of $\tilde t_{c2} \approx 0.043$ estimated from
the linear extrapolation of the previous figure (see the linear extrapolation of the 3$\times$2 cluster data in the previous figure).
}
\end{figure}

To lend support to the validity of this approach, and to demonstrate another advantage of using MSMFT, we have examined and quantified
the spatial correlations present in our multi-site ground state wave functions -- of course, this information is not available when a  site-decoupled single-site theory is employed. That is, we have calculated
\begin{equation}
C(|\mathbf{r}_i-\mathbf{r}_j|) = \overline{\langle \delta {\hat \phi}^\dagger_i \delta {\hat \phi}_j \rangle}, \;\;\; \delta {\hat \phi}_j \equiv {\hat a}_j - \overline{\langle {\hat a}_j \rangle}
\end{equation}
where the overline corresponds to an average over all pairs of sites. Then this function is fit to a functional form appropriate when the system
is in the vicinity of a phase transition, namely
\begin{equation} \label{eq:correlation}
C(|\mathbf{r}_i-\mathbf{r}_j|) = \frac{\exp \left( -| \mathbf{r}_i-\mathbf{r}_j |/\xi \right)}{| \mathbf{r}_i-\mathbf{r}_j |^\eta}
\end{equation}
(Note that for our 2$\times$2 clusters one can access only the distances of $1$ and $\sqrt{2}$ lattice constants.
However, one has many bonds at each of these distances.)
The case of 2$\times$2 clusters is particularly easy for the analysis, since then the correlation length is given by $\xi=-1/\ln C(1)$.
We find that for $\tilde \mu = 0.4$ the correlation length takes its maximum value near the hopping frequency corresponding to $\tilde t_{c2}^{ms}$,
found from Fig.~\ref{fig:fcsingle}, consistent with our simplistic approach of identifying the Bose glass-to-superfluid phase transition shown in that figure.

We have also evaluated $f_c$ and the corresponding correlation function using larger clusters, namely of size 3$\times$2 sites, and our results are shown in Fig.~\ref{fig:xi}. The implementation of this larger lattice gave us two additional distances of $2$ and $\sqrt{5}$ lattice constants.
Because the computational time grows very rapidly with increasing cluster size, this analysis has been limited only to lattices of 30$\times$30 sites.
We found that the parameters describing asymptotical behaviour of the correlation function (see Eq.~(\ref{eq:correlation})) closely resemble the ones found for 2$\times$2 clusters.
Just as for the smaller clusters, the maximum of the correlation length is found to be consistent with the linear fit of 
the $\tilde f_c$ data for $\tilde \mu = 0.4$, as discussed in the caption.
For $\tilde \mu = 0.1$ the comparison of the phase transition identification between these two methods is not as good, and is possibly due to the fact the variation of the correlation length with hopping is very broad and not peaked.
Therefore, while we caution that our simplistic approximation for identifying the Bose glass-to-superfluid phase transition by a linear extrapolation of the condensate fraction data should be only treated as an approximation, our results on the correlation length, at least for $\tilde \mu = 0.4$, lend support for its validity. Furthermore, if larger clusters can be handled numerically, then the use of a MSMFT and the evaluation of the pair
correlation function could provide new information about the behaviour of the phase coherence in the Bose glass phase, 
a topic that we discuss more below.

As mentioned earlier, our numerical result, and previous work \cite{Torino1,Torino2}, on finite lattices, that the condensate fraction is non-zero in the Bose glass phase, is in contrast to the stochastic
mean-field theory, which applies to systems in the thermodynamic limit, as espoused in Refs.~\cite{SMFT1,SMFT2}. 
Our results in Figs. \ref{fig:fc},\ref{fig:fcsingle}, notably the inset of Fig.~\ref{fig:fc} for $\tilde \mu = 0.1$, make clear that
the curvature of $\tilde f_c$~vs.~$\tilde t$ in the immediate neighbourhood of $\tilde t_{c2}^{ss}$ \emph{does not change as one increases system size}. In stochastic mean-field theory this rounding is changed to a discontinuity of the slope with $\tilde f_c$ reduced to zero
below the critical value $\tilde t_{c2}$. 
As mentioned above, we believe that one can address this apparent disagreement through an analysis of the spatial correlations of the order parameter \cite{Aharony}, something that \emph{requires} a multi-site theory beyond what is contained in Fig.~\ref{fig:xi},  but such a resolution of this difference would require numerical results for much larger clusters.

\begin{figure}[ht!]
\includegraphics{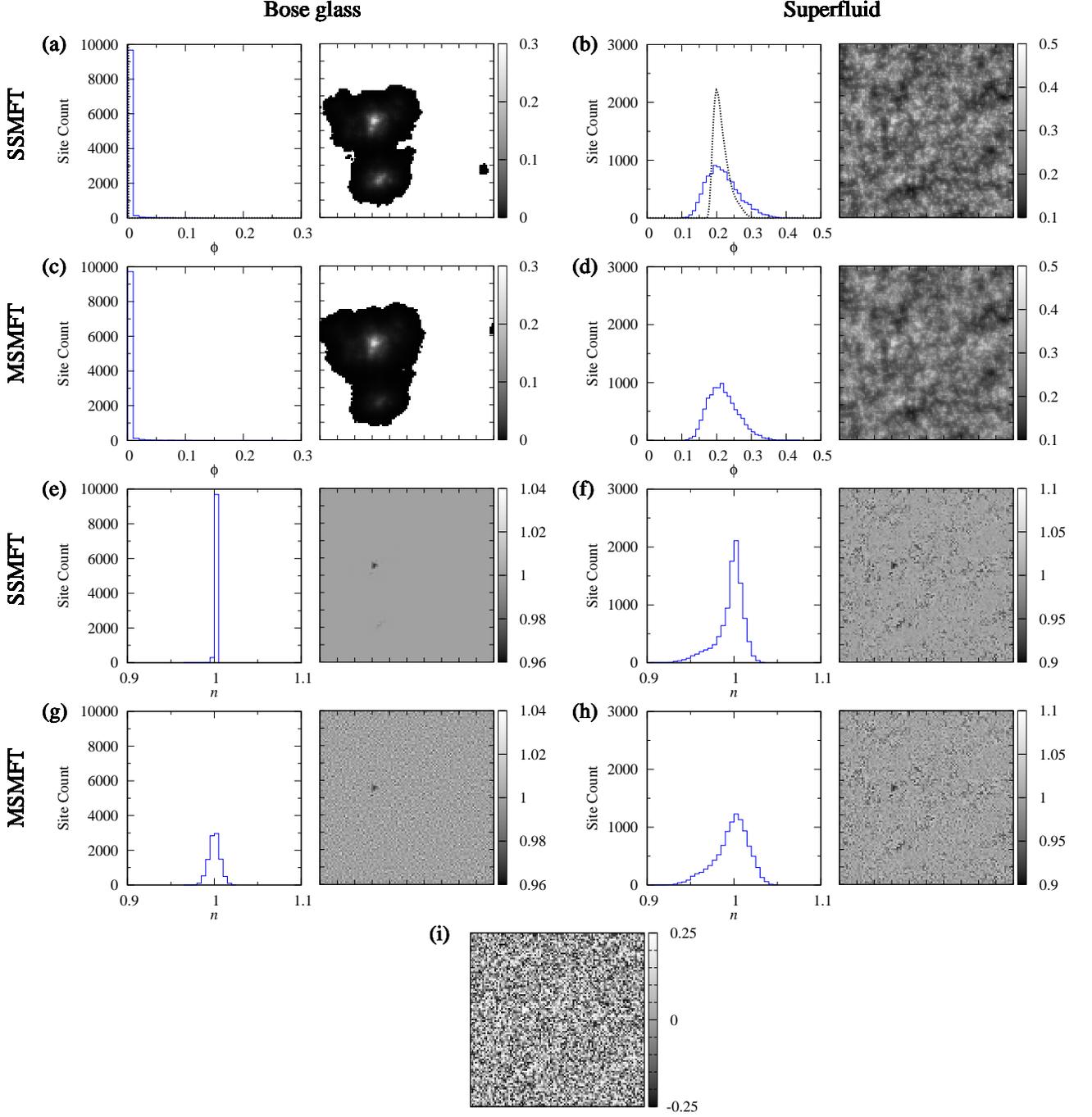}
\caption{ \label{fig:local}
Histograms and position dependencies of the boson densities and order parameters obtained on a lattice of 100$\times$100 sites. In panels (a) and (e) results from single-site mean-field theory (SSMFT) are shown, and in panels (c) and (g) results from multi-site mean-field theory (MSMFT) are shown, all for the system existing in the Bose glass phase. Similarly, in panels (b)
and (f) SSMFT results, and in panels (d) and (h) results for MSMFT, are shown for the system being in the superfluid 
phase.  The distribution of the on-site energies for this particular complexion of disorder is presented in the lower figure (i).  The results have been evaluated for $\tilde W=0.5$ and $\tilde \mu = 0.4$, where the hopping frequencies
are chosen such that in the Bose glass phase $\tilde f_c \approx 10^{-4}$, and in the superfluid phase
$\tilde f_c \approx 0.05$.
The hopping frequencies, corresponding to those condensate fractions, take the following values: 
(a,e) $\tilde t = 0.03721$; (b,f) $\tilde t = 0.04004$; (c,g) $\tilde t = 0.04128$; (d,h) $\tilde t = 0.04397$.
Note that in the Bose glass phase on the majority of the lattice sites the order parameter is vanishingly small, smaller than the convergence criterion, and those regions have been denoted with blank (white) colour.
The normalized results for the $\phi$ probability density function from stochastic mean-field theory (dotted line) are also shown in (a,b).
}
\end{figure}

In Fig.~\ref{fig:local} we show histograms and position dependencies of both 
the order parameters (a-d) and site's occupation numbers (e-h)
in the Bose glass (left panel) and superfluid (right panel) phases obtained from 
the single- and multi-site formulations of mean-field theory. Single-site results are shown in figures a, b, e, f, and
multi-site results in figures c, d, g, h.
For completeness, we also show the random on-site energies $\varepsilon_i$ for the analyzed 100$\times$100 lattice (Fig.~\ref{fig:local}i).
Shades of grey reflect the magnitude of the analyzed quantities (for values, see gradation scales next to each figure).
We choose the values of the hopping integral to correspond to condensate fractions
of $\tilde f_c \approx 10^{-4}$ for the Bose glass phase, and to $\tilde f_c \approx 0.05$ for the superfluid phase.

We find that the maps of the order parameters (Fig.~\ref{fig:local}b,d), as well as the site's occupation numbers (Fig.~\ref{fig:local}f,h), in the superfluid phase for the single-site and multi-site mean-field theories are quite similar. This suggests the benefits of using the simpler single-site theory when studying the superfluid phase (apart from the inability of this formulation to give information about the correlation length).

The situation is more interesting in the Bose glass phase.
In Figs.~\ref{fig:local}a,c we show the position dependencies of the order parameters.
From these maps one can see that even though for the majority of sites the order parameters are equal to zero,
some sites can still develop a non vanishing order parameter.
(Of course, it is this behaviour which leads to the non-zero condensate fraction in the Bose glass phase.)
The sites for which this quantity is smaller than the convergence criterion (in our numerics, this
corresponds to sites with $\phi_i<10^{-4}$) have been denoted as blank (white).
The results are similar for both single- and multi-site formulations, although the multi-site formulation leads
to order parameters which change less abruptly (smoother edges), something that is not unexpected.
In contrast to this, the local boson densities obtained within the single-site mean-field theory are almost
uniformly distributed (and equal to one) over all lattice sites, whereas for the multi-site theory the densities show much greater variation.
Again, this is what one would expect. For example,  
when we look at the results from the multi-site mean-field theory we see that the inclusion of intra-cluster hopping
leads to nonuniform occupation of all sites (in general) for isolated clusters, since, in the areas where the order 
parameter has not been developed, bosons can only hop between sites belonging to a single cluster while 
inter-cluster hopping is suppressed.
To minimize the ground state energy particles need to reorganize themselves, and since the on-site energy, $\varepsilon_i$, differs from site to site, so does the particle density.
In the areas where the order parameter is non-zero bosons can hop between clusters, and this allows them to move outside of regions with high $\varepsilon_i$
(\emph{e.g.}, examine the results corresponding to site (32,56) in Fig.~\ref{fig:local}i).
For the single-site theory bosons can move around the lattice only when the order parameter is non-zero, and this leads to uniform distribution of particles where $\phi_i=0$.
In the superfluid phase the inter-cluster hopping is allowed on the entire lattice (the order parameters are always non-zero).
Therefore the inhomogeneity of the local bosons densities is even stronger --
to minimize the ground state energy, bosons are no longer restrained to particular cluster and can hop around the entire lattice.
Nevertheless, there are areas where the order parameter is small, and these areas
for the single-site mean-field theory show a much weaker variation of the particle density than for the multi-site formulation, simply because the effective hopping probability, $t\phi_i$, is suppressed.
This is reflected in a much narrower local boson density histogram for SSMFT than for MSMFT.

For completeness, in the same figure we also show the order parameter probability distribution obtained within the stochastic (single-site) mean-field theory, discussed in the introduction \cite{SMFT1,SMFT2,SMFTthesis} -- this data corresponds to the dotted lines in  Fig.~\ref{fig:local}b, normalized to allow for a direct comparison of these results with SSMFT.
We found that even though the distributions agree deep inside the superfluid region of the phase diagram
(for example, for $\tilde \mu = 0.4$ one needs the hopping frequency of $\tilde t \gtrsim 0.045$),
in the vicinity of the Bose glass-to-superfluid phase transition they are qualitatively different.
The standard deviation of probability distribution obtained within the stochastic mean-field theory
is much smaller, in this region, than the one we get from the single-site mean-field theory solved on a finite lattice. This is
expected, since within the stochastic theory the order parameter vanishes in the Bose glass phase.

\begin{figure}[ht!]
\includegraphics[scale=0.9]{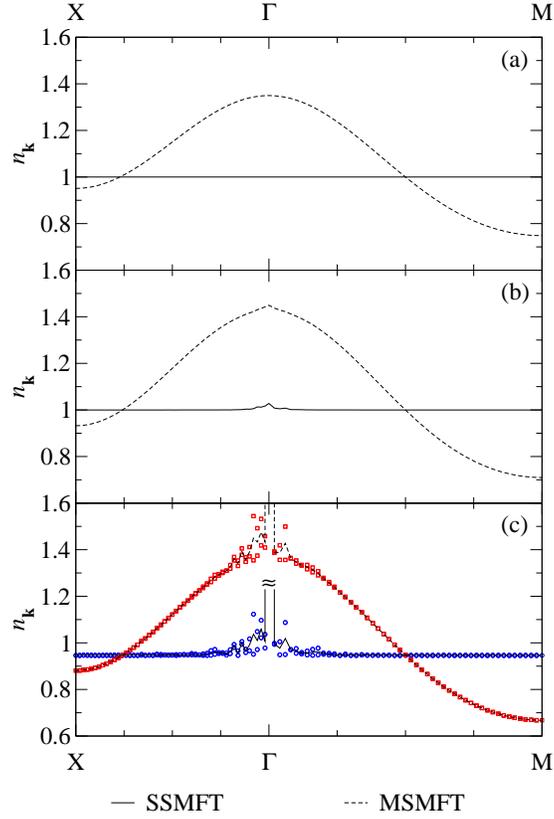}
\caption{ \label{fig:nk}
Momentum distribution of bosons in the first Brillouin zone, along X-$\Gamma$-M path, obtained from the single- and multi-site mean-field theories for $\tilde W = 0.5$ and $\tilde \mu =0.4$
on a 100$\times$100 lattice.
Figure (a) shows the results in the Mott insulator phase, where $\tilde t = 0.035$;
(b) in the Bose glass phase where $\tilde t = 0.03721$ for the single-site and $\tilde t = 0.04128$ for the multi-site formulation;
(c) in the superfluid phase for $\tilde t = 0.04004$ and $\tilde t = 0.04397$, respectively. (These choices are discussed in the text.)
The solid/dashed lines denote the average over corresponding directions of the momentum wave-vector and the dots show actual data points for SSMFT/MSMFT. 
In (c) the original data points, before averaging, are shown, where (blue/red) open circles/squares refer to SSMFT/MSMFT. 
Also, in (c) it should be noted that the $k=0$ data is truncated, since it corresponds to $n_0 \sim 475-480$, and as discussed in the text this corresponds to
the macroscopic occupation of the ground state in the superfluid phase.
(Note the reduction of $n_k$ for the SSMFT from 1 to 0.95 away from $k=0$, and a similar reduction for MSMFT.)
}
\end{figure}

In Fig.~\ref{fig:nk} we show the momentum distribution of bosons, in all three different phases, given by
\begin{equation} \label{eq:nk}
n_{\mathbf{k}} = \frac{1}{N_s} \sum_{i,j} e^{i \mathbf{k} \cdot (\mathbf{r}_i-\mathbf{r}_j)} \langle \hat{a}_i^\dagger \hat{a}_j \rangle
\end{equation}
The results shown have been obtained for a single complexion of disorder on a 100$\times$100 lattice.
(This is the same realization of disorder potential as the one used for the position space results in Fig.~\ref{fig:local}.)
We found that in the Mott insulator phase (Fig.~\ref{fig:nk}a) the momentum distribution obtained within the single-site theory is uniform -- all states with momenta $\mathbf{k}$ are equally occupied.
Note that in this region of the phase diagram all order parameters are identically equal to zero, which leads to $\langle \hat{a}_i^\dagger \hat{a}_j \rangle = \langle \hat{n}_i \rangle \delta_{ij}$.
Therefore, we find that $n_{\mathbf{k}} = N_b/N_s = n$, which for the set of $\tilde t$, $\tilde \mu$ and $\tilde W$ parameters that we chose, ends up to be $n = 1$.
However, we observe that the momentum distribution resulting from the multi-site formulation has a characteristic bell shape
that is present throughout the entire phase diagram, even in the Bose glass and superfluid phases.
This is a consequence of including the intra-cluster hopping in the evaluation of the eigenstates of the zeroth order Hamiltonian (see Eq.~(\ref{eq:K0-I})),
which, as emphasized above, leads to the multi-site theory retaining some of the inter-particle spatial correlation effects.
Because of this, the momentum distribution of bosons in the Mott insulator phase is not a simple sum of the occupation numbers on each site,
instead containing other terms for which the sum in Eq.~(\ref{eq:nk}) runs over the individual cluster sites.
These effects are completely washed out if the single-site formulation is used.
In the Bose glass phase (Fig.~\ref{fig:nk}b) the disorder starts to play an important role, as bosons can scatter off of the disorder potential,
which results in varying particle numbers being present around the $\mathbf{k}=0$ momentum.
As one expects, the presence of disorder breaks all symmetries of this distribution; however, the inversion symmetry stays intact (not shown).
A further increase of the hopping frequency causes the system to go into the superfluid phase, and therefore the 
ground state starts to be macroscopically occupied, which is reflected in the momentum distribution by a peak near
the $\mathbf{k}=0$ momentum (Fig.~\ref{fig:nk}c).
In the figure, this leads to $n_0$ being around 475-480 (truncated in the figure for clarity), and which is compensated
by the reduction of $n_{\bf k}$ at other momenta such that the boson densities sums are similar in all panels.  

Taken together, the above results makes clear that it is in the Bose glass phase that the advantages of using MSMFT instead of SSMFT are most noticeable. However, as seen in the last figure, and in our discussion of the correlation length,
there are quantities for which the MSMFT provides new information.


\section{Discussion}

We have derived a multi-site mean-field theory for the disordered Bose-Hubbard model, and compared the results found from such an approach with those obtained from the more familiar site-decoupled single-site theory. The main differences between the single-site and multi-site formulations of mean-field theory are:
(I) One includes the effects of intra-cluster hopping, thereby producing a perturbation theory that includes intra-cluster particle-particle correlation effects
in the unperturbed eigenstates (see Eq.~(\ref{eq:K0-I})).
(II) One incorporates some aspects of the spatial variation of the on-site energies in the unperturbed eigenstates (the
cluster eigenstates of Eq.~(\ref{eq:K0eigs}) depend on intra-cluster complexions of disorder),
while in single-site theory for the unperturbed eigenstates there is only one on-site energy.

We found that the multi-site theory predicts a stable Mott insulator phase for greater hopping frequencies than the single-site formulation.
This is consistent with parallel studies of the similar problem for homogeneous lattices \cite{Tom}.
Additionally, for both SSMFT and MSMFT theories we showed that the stability of the Mott insulator phase depends on the complexion of disorder --
even for lattices of size 200$\times$200 sites our results did not converge to a single value of $\tilde t_{c1}$, the Mott
insulator to Bose glass transition. It would be interesting to see if this dependence on the complexion of disorder is also observed in the experimental systems.
(Note that it is straightforward to include the effects of a harmonic trapping potential into our MSMFT numerics.)
Of course, we are limited to finite lattices, so no rare Lifshitz regions can be
included, and other theoretical approaches are more appropriately suited for examining their role in determining
the phase diagram of the system in the thermodynamic limit. This limit may also lead to the results of stochastic MFT
theory being correct for the condensate fraction, namely that it vanishes in the Bose glass phase, although such a conclusion would need to be shown to be robust under the extrapolation of this inventive approach to a multi-site mean-field theory.

We found that in the Bose glass phase the position distribution of particles obtained within multi-site theory is correlated with the underlying random potential of the on-site energies, whereas
the single-site theory predicts that the vast majority of sites are uniformly occupied with a given number of bosons per site
(only sites which developed a non-zero order parameter have varying occupation numbers).
This is clearly a result of incorporating the spatial variation of the on-site energies in the eigenstates of each isolated
cluster.

We have also shown that these two formulations of the mean-field theory give rise to two qualitatively different momentum distributions of particles.
The multi-site formulation predicts a characteristic bell shape thereof,
a feature that is clearly a product of the inclusion of the intra-cluster hopping in the zero order perturbation Hamiltonian (see Eq.~(\ref{eq:nk})).
The single-site theory, however, does not show such a behaviour since it does not incorporate any of the particle-particle correlations effects.

We also presented a way of identifying -- approximately -- the Bose glass-to-superfluid phase transition
and therefore completing full phase diagram of the Bose-Hubbard model, solved on a finite size lattice for a
given complexion of disorder, from the condensate fraction alone. By examining the correlation length we have seen that this approximation is expected
to be accurate near the tip of Mott lobe, but less accurate away from it.

Taken together, these results suggest that the MSMFT better accounts for the variations of the boson density and order parameter distribution that are expected in the Bose glass phase, and it seems here that the potential utility of this approach could best be exploited. Future work
will discuss the dynamics that are found, in the Bose glass phase, using MSMFT. The differences between
the two formulations in the superfluid phase do not seem to be as dramatic, although quantities such
as the correlation length and the momentum distribution are accessible via MSMFT.

\acknowledgements

We acknowledge helpful conversations with Tom McIntosh and Eugene Zaremba on the development
of the multi-site mean-field theory for ordered lattices. One of us (RJG) thanks Andrew Parks for discussions
regarding algorithms for multidimensional numerical integrations. This work was supported in part by the NSERC of Canada.

\newpage

\bibliographystyle{unsrt}

\end{document}